\documentclass[aps,rmp,groupedaddress]{revtex4}

\newif\ifpdf
\ifx\pdfoutput\undefined
\pdffalse 
\else
\pdfoutput=1 
\pdftrue
\fi

\ifpdf
\usepackage[pdftex]{graphicx}
\else
\usepackage{graphicx}
\fi
\usepackage{hyperref}

\usepackage{color}

\begin{document}

\ifpdf
\DeclareGraphicsExtensions{.pdf, .jpg}
\else
\DeclareGraphicsExtensions{.eps, .jpg}
\fi

\def\hslash{\hbar}
\def\imag{i}
\def\grad{\vec{\nabla}}
\def\div{\vec{\nabla}\cdot}
\def\curl{\vec{\nabla}\times}
\def\DDt{\frac{d}{dt}}
\def\ddt{\frac{\partial}{\partial t}}
\def\ddx{\frac{\partial}{\partial x}}
\def\ddy{\frac{\partial}{\partial y}}
\def\lap{\nabla^{2}}
\def\divv{\vec{\nabla}\cdot\vec{v}}
\def\gradS{\vec{\nabla}S}
\def\vvec{\vec{v}}
\def\wc{\omega_{c}}
\def\<{\langle}
\def\>{\rangle}
\def\Tr{{\rm Tr}}
\def\Csch{{\rm csch}}
\def\Coth{{\rm coth}}
\def\Tanh{{\rm tanh}}
\def\g2{g^{(2)}}

\preprint{IJQC/2003 Sanabel Symposium}

\title{Energy relaxation dynamics and universal scaling laws in organic light emitting diodes}
\author{Eric R. Bittner}
\email[email:]{bittner@uh.edu}
\affiliation{Department of Chemistry, University of Houston, Houston, TX 77204-5003}
\author{Stoyan Karabunarliev}
\email[email:]{karabuna@uh.edu}
\affiliation{Department of Chemistry, University of Houston, Houston, TX 77204-5003}

\date{\today \,\,\, For IJQC-Sanibel Symposium}

\begin{abstract}
Electron-hole (e-h) capture in luminescent conjugated polymers (LCPs) is
modeled by the dissipative dynamics of a multilevel electronic system
coupled to a phonon bath. Electroinjected e-h pairs are simulated by a
mixed quantum state, which relaxes via phonon-driven internal
conversions to low-lying charge-transfer (CT) and excitonic (XT) states.
The underlying two-band polymer model reflects PPV and spans monoexcited
configuration interaction singlets (S) and triplets (T), coupled to
Franck-Condon active C=C stretches and ring-torsions.  Focusing entirely
upon long PPV chains, we consider the recombination kinetics of an
initially separated CT pair.  Our model calculations indicated that S
and T recombination proceeds according to a branched, two-step mechanism
dictated by near e-h symmetry.  The initial relaxation occurs rapidly
with nearly half of the population going into excitons ($S_{XT}$ or $T_{XT}$),
while the remaining portion remains locked in metastable CT states. 
While formation rates of $S_{CT}$ and $T_{CT}$ are nearly equal, $S_{XT}$ is formed
about twice as fast  $T_{XT}$ in concurrence with experimental observations of
these systems.  Furthermore, breaking e-h symmetry suppresses the XT to
CT branching ratio for triplets and opens a slow CT$\rightarrow$ XT conversion
channel exclusively for singlets due to dipole-dipole interactions
between geminate and non-geminate configurations.   Finally, our calculations 
yield a remarkable linear relation between chain length and singlet/triplet branching ratio
which can be explained in terms of the binding energies of the respective final excitonic states
and the scaling of singlet-triplet energy gap with chain length. 
\end{abstract}
\pacs{}
\maketitle

\section{Introduction}
Since the discovery of electroluminescence (EL) in poly(p-phenylene
vinylene) PPV,\cite{ref1} considerable efforts have gone in development and
commercialization of light-emitting diodes (LEDs), based on luminescent
conjugated polymers (LCPs). While much progress has been made in
improving emission efficiency and color control,\cite{ref2} the detailed physics
of charge transport and light generation in these materials is
relatively poorly understood in comparison with inorganic
semiconductors. The one-dimensional delocalization of electrons and
substantial electron-hole and electron-lattice interactions render the
delineation of neutral and charged excitations in LCPs very complex and
demanding.  Moreover, a dynamical description is needed in order to
capture the transient relaxation processes that occur when charged
species approach each other, recombine and decay.

Photogenerated singlet e-h pairs in LCPs typically relax to highly
emissive $S_1$ excitons prior to decay. In contrast, electroinjected
electrons $(e^-)$ and holes $(h^+)$ are not spin correlated and can
combine to form both singlet and triplet excitons.
 \begin{eqnarray} e^-
+ h^+ \begin{array}{lll} \stackrel{k_{S}}{\longrightarrow} & S_1
&\stackrel{k_{rad}}{\longrightarrow} S_o + h\nu \\
\stackrel{k_{T}}{\longrightarrow} & T_1 &
\stackrel{k_{so}}{\longrightarrow} S_o  \\ \end{array}
 \end{eqnarray}
Because the decay of the triplet exciton $T_1$ is nearly radiationless
and relaxation to $S_o$ occurs via spin-orbit coupling, quantum
efficiency in electroluminescence $(\eta_{EL})$ is only a fraction of
that for photoluminescence $(\eta_{PL})$. If the rate of e-h capture is
spin-independent,  $S_1 $ and $T_1$ excitons are formed in a 1:3 ratio
as dictated by spin degeneracy. In such a case, 75\% of the bound e-h
pairs are non-emissive leading to a theoretical maximum of $\eta_{EL} =
0.25\,\eta_{PL}$. Nonetheless, efficiencies of up to $\eta_{EL} = 0.50
\, \eta_{PL}$ have been achieved independently on PPV-based LEDs in the
laboratories of Heeger\cite{ref2} and Friend\cite{ref3}. From this it
has been inferred that in organic light-emitting polymers, singlet $e-h$
 capture is intrinsically more efficient than the respective triplet
process $(k_S > k_T)$.

If we assume that the formation kinetics is first order such that the
singlet and triplet exciton population formation rates are proportional
to the total population of charge-transfer (CT) states in the system,
then the formation cross-section \({{\sigma }_{s  }}\)for singlet
excitons as given by the ratio of singlets to total excitons by
\begin{eqnarray} {{\sigma }_s}=\frac{{{  {N}}_s}(t)}{{{  {N}}_s}(t)+3{{ 
{N}}_T}(t)}=\frac{r}{r+3  } \end{eqnarray} where
\(r={k_S}/{k_T}=\sigma_S/\sigma_T\). In order to obtain the nearly 64\%
EL efficiencies reported by various groups\cite{ref2,ref3}
in various organic polymer LED devices,  
$r$ must be in the range of $3 < r < 5$.  
Such ratios were also obtained for a wide range of organic LCPs  
via photoabsorption/detection of magnetic resonance (PADMR) 
experiments by Wohlgenannt, {\em at  al} \cite{ref5a,ref5b} 
which measures the population of photo-generated intrachain parallel and anti-parrallel
polaron pairs in the presence of a magnetic field.  This later work\cite{ref5b} is
particularly significant in that it establishes a universal 
linear scaling relationship between $r$ and inverse conjugation length,
which holds for a wide range of LCP systems.   

Various mechanisms favoring the formation of $S_1$ versus $T_1$ have been
proposed for both interchain and intrachain e-h collisions.  Using
Fermi's golden rule, Shuai, Bredas et al.\cite{ref6,ref6b} indicate that the $S$ cross
section for interchain recombination can be higher than the triplet one
due to bond-charge correlations.  Wohlgenannt {\em et al.}\cite{ref5a} resort to a
similar model of two parallel polyene chains.    Both of these works
neglect vibronic and relaxation effects.  In simulating the intrachain
collision of opposite polarons, Kobrak and Bittner \cite{ref7,ref7b} show that formation
of $S_1$ is enhanced by the near-resonance with the free e-h pair.  The
result reflects the fact that spin-exchange renders $T_1$ more tightly
bound than $S_1$,\cite{ref8,ref8b} and hence more electronic energy must be dissipated 
by the phonons in
the formation of the former. The energy-conservation constraints in
spin-dependent e-h recombination have been analyzed by Burin and Ratner\cite{ref9}
in an essential-state model. The authors point out that nonradiative
processes (internal conversion, intersystem crossing) must entail C=C
stretching vibrons since these modes couple most strongly to $\pi\rightarrow\pi^*$
excitations.   Recent work by Tandon et al. suggest that irrespective of the 
recombination process, interchain or intrachain, the direct transition to 
form singlets should always be easier than triplets due to its smaller binding energy
relative to the triplet.~\cite{Tandon}.  
Lastly, a comprehensive review of this work detailing the experiments and summarizing the 
theory of this effect is forthcoming.\cite{MWreview}

In this work, we describe e-h capture in LCPs in terms of the
dissipative dynamics of a multi-level electronic system coupled to a phonon bath.  
Our model is based upon a two-band polymer model introduced by Soos, Mazumdar, {\em et al}
\cite{ref10,ref10b}
augmented by the coupling of electronic excitations to a bath of vibrational modes. 
First, we  review a time-dependent 
model for simulating via quantum chemical methods the 
transient relaxation dynamics of an excited state in an extended system.\cite{SKEB1,SKEB3}
The methodology we develop herein 
is applicable to both excitonic transfer (i.e. F{\"o}rster) as well as charge-transfer
states in a general conjugated polymer system.
At the present, we restrict our attention to mono-excited closed-shell 
systems;  however, the approach can be extended to the general case for 
radicals.   
Secondly, we examine a specific physical process involving the 
collision and relaxation of injected electron-hole pairs in luminescent 
conjugated polymers, such as PPV.\cite{SKEB2}  Our calculations underscore the role that
vibronic coupling and electron-hole symmetry play in determining the 
singlet-triplet branching ratio  as the CT state relaxes to form bound electron/hole
excitonic pairs.   Finally, 
we explore the relation between chain length and the singlet/triple branching ratio. 
We show that our methodology reproduces the remarkably universal linear relation 
between conjugation length and $r$ as evident in a wide range of conjugated polymer
materials.  To our knowledge, ours is the first molecular based model which accurately 
predicts this linear relation for a specific molecular system and provides a 
rationalization for this trend based upon the variation of the exchange energy with 
increasing chain length. 

\section{Methodology}

\subsection{Two-band polymer model + phonons} 
The Hamiltonian for the coupled system is
\begin{eqnarray}
H &=& H_{el} + H_{ph} + H_{el-ph}\nonumber \\
&=& \sum_{mn}(F^o_{mn}|{\bf m}\>\<{\bf n}|  \nonumber \\
&+&\frac{1}{2}\sum_{\mu\nu}\kappa_{\mu-\nu}q_\mu q_\nu  + \delta_{\mu\nu}p_\mu^2 \nonumber \\
&+& \sum_{mn\mu}\left( \frac{\partial F_{mn}}{\partial q_\mu}\right)_o
q_\mu |{\bf m}\>\<{\bf n}| 
\end{eqnarray}
The electronic Hamiltonian $H_{el}$  represents the configuration interaction of 
localized singlet of triplet configurations $|{\bf m}\>=| \overline{m}m\>$ with a valance hole in repeat unit
$\overline{m}$ and a conduction electron in repeat unit $m$.  
For example,  the $|\overline{m}m\>$ geminate monoexcited configuration for a 6 repeat-unit system  is 
\def\ud{\underline{\uparrow\downarrow}}
\def\unocc{\underline{\,\,} }
\def\ua{\underline{\,\uparrow}}
\def\da{\underline{\downarrow\,}}
\begin{eqnarray}
|\overline{m}m\>=
\left(
\odot-\odot-\otimes-\odot-\odot-\odot
\right)
\end{eqnarray}
where as 
\begin{eqnarray}
|\overline{m}n\>=
\left(
\odot-\oplus-\odot-\odot-\ominus-\odot
\right)
\end{eqnarray}
represents a charge-transfer configuration.  Our notation is such that $(-\odot-)$ represents a neutral site in the 
ground-state configuration, $(-\oplus-)$ denotes a site with a hole, $(-\ominus-)$ a negatively charged site with an 
excess electron, and $(-\otimes-)$ a neutral excitonic site. 

System specific 
information is incorporated by using valance and conduction band Wannier functions (WF) $|\overline{m}\>$ and
$|m\>$ as the single particle basis.  
For the single-particle band-structure
 we use a Hartree-Forck Pariser-Parr-Pople (PPP) approximation which can be 
parameterized easily for a variety of luminescent conjugated polymer systems.  
and construct maximally optimized Wannier functions.
The single-particle terms, $F_{mn}^o$, are 
derived at the ground state equilibrium configuration, $q_\mu = 0$, from the Fourier components $f_r^o$ and 
$\overline{f}_r^o$ of the 
band energies in pseudomomentum space.
\begin{eqnarray}
F_{mn}^o &=& \delta_{\overline{m}\overline{n}}\<m| f^o|n\> - \delta_{mn}\<\overline{m}|\overline{f}^o|\overline{n}\>\nonumber \\
&=&\delta_{\overline{m}\overline{n}}f_{m-n}^o - \delta_{mn}\overline{f}_{\overline{m}-\overline{n}}^o
\end{eqnarray}
For extended $\pi$-systems with electron-hole symmetry, the localized electron/hole energy levels and transfer
parameters are related by $\overline{f}_r = -f_r$ and reflect the 
cosine-shaped valance and conduction bands of half-width $f_1^o$ and centered at
$\pm f_o^o$ respectively.    

The two-particle interactions are spin dependent with 
\begin{eqnarray}
V_{mn}^T = -\<m\overline{n}||n\overline{m}\> \\
V_{mn}^S = V_{mn}^T  + 2 \<m\overline{n}||\overline{m}n\>.
\end{eqnarray}
with 
$$
\<m\overline{n}||i\overline{j}\> = \int d1\int d2 \phi_m(1)^*\phi_{\overline{n}}^*(2) v(12)\phi_i(1)\phi_{\overline{j}}(2)$$
With the exception of 
geminate WFs, orbital overlap is small such that the two-body interactions 
are limited to Coulomb and exchange integrals reflecting e-h attraction and 
spin-exchange coupling non-geminate configurations 
such as 
\begin{eqnarray}
J_{|m-n|}=\langle m\overline{n}||m\overline{n}\rangle = J_o(1-r/r_o)^{-1}\\
K_{|m-n|}=\langle m\overline{n}||\overline{n}m\rangle = K_o\exp(-r/r_o)
\end{eqnarray}
and transition dipole-dipole integrals coupling only geminate singlet 
electron-hole pairs,
\begin{eqnarray}
D_{|m-n|}=D_o(r/r_o)^3.
\end{eqnarray}
In the absence of e-h symmetry, dipole-dipole coupling between 
singlet geminate and non-geminate configurations is introduced through this term. 
As we shall explore later in this paper, this opens a channel for coupling 
singlet charge-transfer states  to singlet excitonic states but does not allow 
for coupling between the triplet CT to XT states. 

Finally, the electron-phonon coupling term, $H_{el-ph}$ describes the local variation
of the single-particle band gap and is taken to be linear a set of phonon 
normal-mode coordinates $q_\mu$  which are localized within a single unit-cell. 
The linear coupling strength for both the electron and holes, $(\partial f/\partial q_\mu)_o
= - (\partial \overline{f}/\partial q_\mu)_o$, is adjusted empirically 
as to reproduce the the vibronic progressions observed in the single-photon absorption
and emission spectra.   Correspondingly, the vibrational term, $H_{ph}$ models 
weakly dispersec optical phonon branches in the frequency ranges of the dominant 
Franck-Condon active modes observed in the experimental spectra.  
For the case of PPV considered here, the phonon term  consists of two
sets of local harmonic oscillators with weak nearest-neighbor coupling,
representing two dispersed optical phonon branches, centered at 1600 and
100cm$^{-1}$, respectively. These frequencies roughly correspond to C=C bond
stretches and ring-torsional motions (librations), which dominate the
Franck-Condon activity in PPV and related polymers.\cite{ref14,ref14b,ref14c} 
Huang-Rhys factors (dimensionless el-ph coupling constants) for $S_1$ 
of the isolated monomer were subsequently set to 0.8 for the high-frequency 
mode and 10 for the low-frequency mode.

\subsection{Diabatic Representation}

\begin{figure}[t]
\includegraphics[width=6.0in]{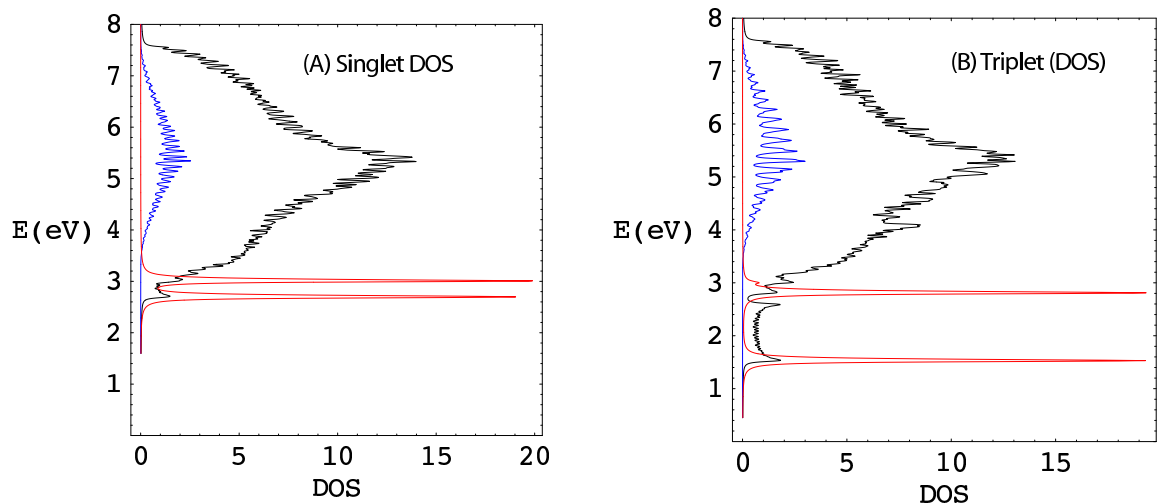}
\caption{Density and population of singlets (A) and triplets (B): density of states 
- black; initial population - blue; population at 100ps - red. }\label{DOS-PPV}
\end{figure}

\begin{figure}[b]
\includegraphics[width=6.0in]{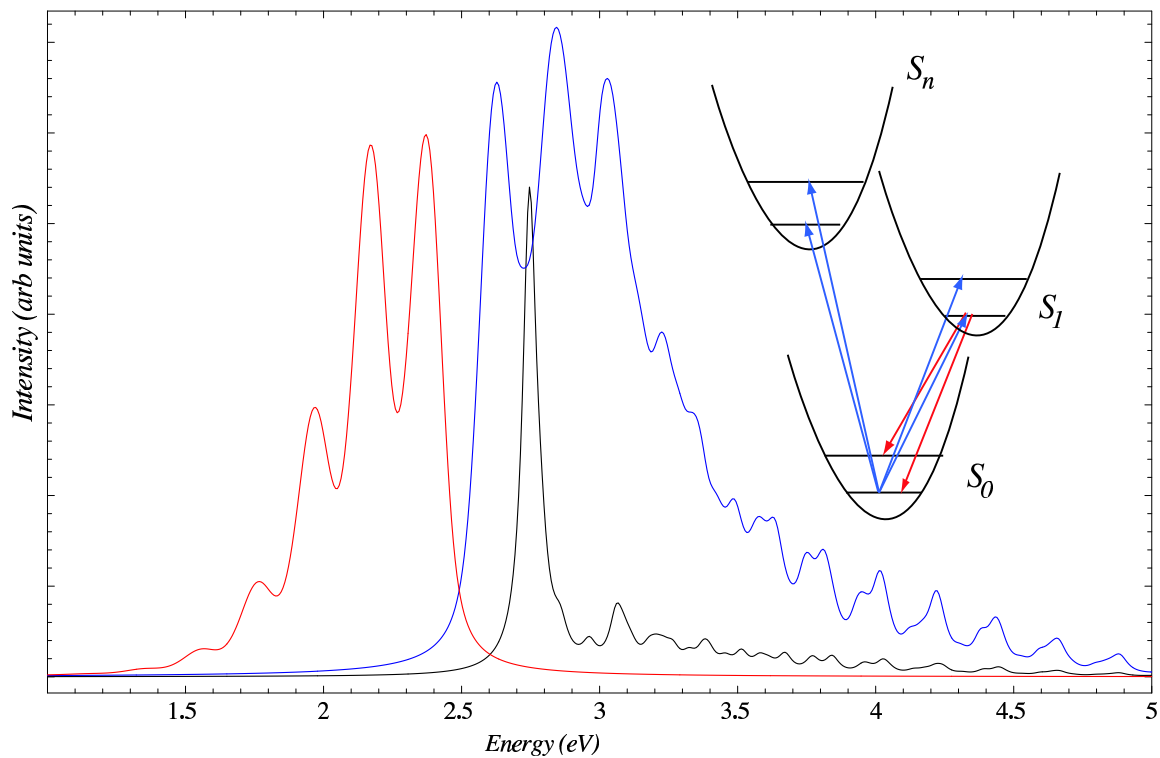}
\caption{Computed Condon spectral densities for $(PPV)_{32}$. 
Absorption (blue) and red-shifted emission (red)
spectra between the adiabatic ground state and the
excited states and the black curve is the oscillator strength for vertical excitation
from the  ground state.   Spectral features correspond to excitation of
 C=C vibronic stretching modes in the excited state (upon absorption)
 and in the ground state (upon emission) as detailed in the inset figure. }\label{spect}
\end{figure}

Having set up the primitive Hamiltonian and parameterized it according to a
model physical system, we can obtain the diabatic electronic/vibrational structure
by separately diagonalizing $H_{el}$ and $H_{ph}$ yielding a series of 
vertical excited states with energies $\varepsilon_a^o$
 (from $q_\mu = 0$) and a normal modes with frequencies $\omega_xi$.  Written in 
the diabatic representation
\begin{eqnarray}
\tilde{H} &=& \sum_a \varepsilon_a^o |{\bf a}\>\<{\bf a}| \nonumber \\
& +& \frac{1}{2}\sum_{\xi}\left( \omega_\xi^2 Q_{\xi}^2 + P_\xi^2\right)\nonumber \\
& +& \sum_{a b \xi} g_{ab\xi}^o Q_\xi |{\bf a}\>\<{\bf b}| .\label{Hdiabatic}
\end{eqnarray}
The diabatic coupling term $g_{ab\xi}^o$ is simply the original electron-phonon coupling 
tensor transformed into this new representation. 
The density of states (DOS) for vertical singlet and triplet excitations 
from the ground state for a PPV model with 32 repeat units ($PPV_{32}$) is shown in 
Fig.~\ref{DOS-PPV}, as parameterized from the $\pi$-band structure and
Wannier functions of the extended polymer.   
Although the Stokes shifts from adiabatic relaxation of the excited states
is missing in this calculation, the lowest singlet $S_1$ and triplet $T_1$ 
excitons differ significantly in terms of their binding energies.  Here, 
$T_1$ is roughly 1eV lower than the $S_1$ in general agreement with 
experimental and theoretical estimates.   We can also use the diabatic representation 
to calculate the relaxed excited states and their respective potential energy 
curves.  This we achieve by variational minimization of the adiabatic energy
$\varepsilon_a = \<{\bf a}|\tilde{H}|{\bf a}\>$  of each individual state
according to 
\begin{eqnarray}
\frac{\partial \varepsilon_a}{\partial Q_\xi} = g_{aa\xi} + \omega_\xi^2 Q_\xi = 0.
\end{eqnarray}
The relaxed excited states are eigenstates of $\tilde{H}_{el} + \tilde{H}_{ph}$ at 
state specific conformations, $Q_{\xi,a} \ne 0$, displaced from the ground state
at $Q_\xi = 0$.  Such displacements modify the optical spectra by moving states into and out
of regions of strong Franck-Condon coupling to the ground state. ~\cite{ref15,ref15b}
  Thus, we can 
test the parameterization of the model by computing the Condon spectral 
density for single photon absorption.  The cumulative spectral density 
of $PPV_{32}$ is given in Fig.~\ref{spect} for both the vertical
and Condon approximation.   
Individual electron-vibrational
transitions are taken to give Lorentzian lineshapes of half-width
0.01eV.   The oscillator strengths for the $S_o\rightarrow S_a$ 
vertical transitions were computed by assuming that only geminate and 
nearest neighbor e-h pairs are dipole coupled to $S_o$.  Since $S_1$  is 
approximately and all-symmetric combination of such configurations, the bulk of the 
spectra density is concentrated in this state.   Furthermore, the 
$S_o\rightarrow S_1$ vibronic band is peaked in the $0\rightarrow 1$ C=C
stretching feature in agreement with the experimental spectra.   
Vibronic structure from librations is smeared and
produces spectral broadening proportional to Huang-Rhys factors times
phonon frequencies. The low-frequency coupling also determines the
$\approx$ 0.15eV Stokes shift of the $0\rightarrow 0$ C=C stretch feature from the adiabatic
origin of $S_1$ at 2.34eV.

\subsection{Relaxation dynamics}

Separated CT states 
are not eigenstates of the diabatic Hamiltonian and will evolve in time 
according to the time-dependent Schr\"odinger equation. 
Unfortunately, a complete description of the vibronic 
dynamics is unfeasible due to the enormous size of the state-space.
However, we can consider the phonons as a dissipative finite-temperature
bath for the electrons and derive reduced equations of motion for the 
electronic dynamics.  For this we turn to the Liouville-von Neumann equation
for the evolution of the electronic density matrix
\begin{eqnarray}
i\hbar \dot \rho = [\tilde{H}_{el},\rho] + \frac{i}{\hbar}{\cal R}\rho
\end{eqnarray}
where the first term represents the unitary evolution of the uncoupled electronic states in the 
diabatic representation and the second term the non-unitary, dissipative 
dissipative dynamics due to the coupling to the phonon bath.  
Since the electronic state-space consists of  roughly 100 energy levels, we
restrict our attention to the population dynamics and decouple
populations from coherences according to the Bloch model,
\begin{eqnarray}
{\cal R}_{aabb} = -k_{ab} + \delta_{ab}\sum_c k_{ac}\\
{\cal R}_{abab} = \frac{1}{2}\sum_c (k_{ac}-k_{bc})
\end{eqnarray}
where $k_{ab}$ are the rates of elementary interstate transitions. 
For internal conversions within the diabatic excited states, we assume that the 
phonons thermalize rapidly on the time-scale of the electronic dynamics
such that the one-phonon transitions rates can be determined within the 
Markov approximation
\begin{eqnarray}
k_{ab} = \pi\sum_{\xi} \frac{(g_{ab\xi}^o)^2}{\hbar\omega_\xi} (n_\xi + 1)
(
\Gamma(\omega_\xi-\omega_{ab})-
\Gamma(\omega_\xi+\omega_{ab}))
\end{eqnarray}
Here $n_\xi$ is the Bose-Einstein distribution of phonons at T = 300K and $\Gamma$ is
the empirical broadening of 0.01eV. Note that for an elementary $a\rightarrow b$
population transfer to occur there must a phonon mode $\omega_\xi$ at that transition
frequency $\omega_{ab}$.  Thus, phonon mediated dynamics is restricted to the
excited state manifold where the energy level spacing is commensurate with the 
phonon energies.  In contrast, single-photon processes occur across the $S_{o}\rightarrow S_1$ 
optical gap and the rate of spontaneous emissions is given by 
\begin{eqnarray}
k_{a0} = \frac{|{\bf \mu}_{a0}|^2}{6\epsilon_o\hbar^2}(1+n(\omega_{a0}))\frac{\hbar \omega_{a0}^3}{2\pi c^3}
\end{eqnarray}
where ${\bf \mu}_{a0}$ are the transition dipoles of the excited singlets.  
These we can comput directly from the Wannier functions or empirically 
from the photoluminescence decay rates for a given system.  Photon mediated transitions
between excited states are highly unlikely due to the $\omega_{ab}^3$ density factor of the optical 
field.  In essence, so long as the non-equilibrium vibrational dynamics 
is not a decisive factor, we can use these equations to trace the relaxation 
of an electronic photo- or charge-transfer excitation from its creation to 
its decay including photon outflow measured as luminescence.

\section{Electron-hole capture kinetics}
\subsection{Capture cross-section and chain-length}
\begin{figure}[t]
\includegraphics[width=6.0in]{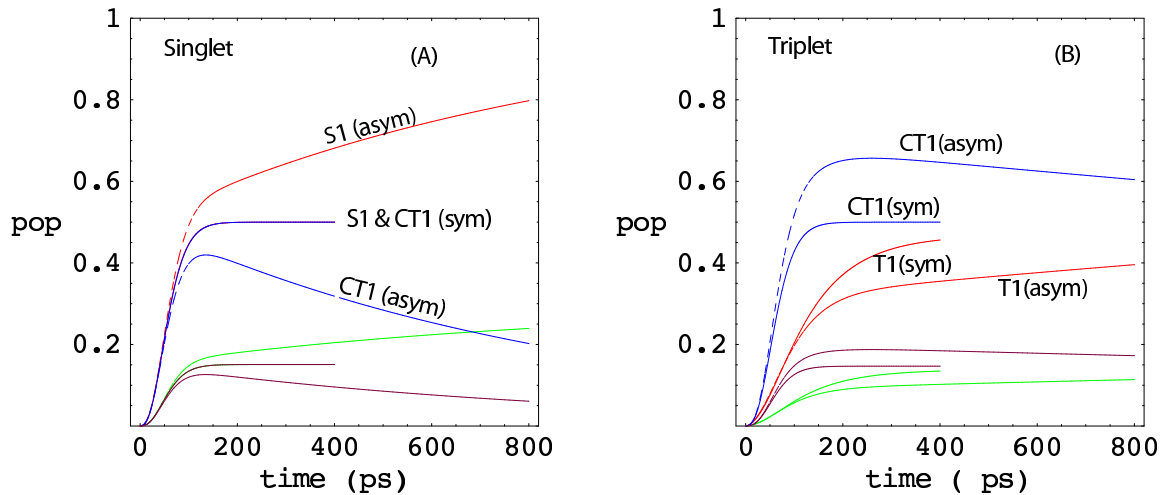}
\caption{E-h capture population dynamics of essential states with and  without e-h symmetry.
for singlet (A) and triplet (B) cases.
Solid curves are with e-h symmetry enforced and dashed curves are for broken e-h symmetry.
In both figures, we show the population dynamics for the 
lowest two excitonic and charge-transfer states.}\label{rates}
\end{figure}

In modeling of e-h capture, propagation starts from the free e-h pair
represented by the charge-transfer (CT) configuration with an electron
and a hole on the far ends of the polymer chain, for example:
\begin{eqnarray}
|\psi_0\>=
\left(
\ominus-\odot-\cdots-\odot-\oplus
\right)\label{CTstate}
\end{eqnarray}
We considered chains with $n$ = 2,4, 8, 16, and 32 repeat units corresponding to 
longer and longer polymer chains.  The salient data from each of these calculations is
presented in Table~\ref{tab1}.  
In Fig.\ref{DOS-PPV}  we show the
diabatic density of states (DOS), and the spectral distribution of the
starting configuration in S and T state-spaces for a 32 repeat unit chain $(PPV)_{32}$. 
 Singlet and triplet
populations are also given at an intermediate stage of relaxation at
100ps. In both cases, the system evolves to a metastable superposition
state, where half of the density remains locked in CT states ($S_{CT}$ or
$T_{CT}$) above the lowest excitons ($S_{XT}$ or $T_{XT}$).\cite{ref17}
 The branching into CT and
XT channels is due to e-h symmetry, which separates excited states into
even and odd representations under e-h transposition. Thus, XT states, which are even,
are not vibronically coupled to CT states, which are odd.  As shown in
Fig.\ref{rates}a, $S_{XT}$ and $T_{XT}$
 are formed at different rates. Whereas $S_{XT}$, $S_{CT}$ and
$T_{CT}$ show almost parallel population growths, formation of $T_{XT}$ is far
slower.

\begin{table}[b]
\caption{Collected data for various chain lengths (symmetric case).  Here, 
$\tau_S$ and $\tau_T$ are
the formation half-times ($\tau = \ln(2)/k$), $r = k_s/k_t$, and $\phi=r/(r+3)$ is the 
electro-luminescent efficiency.}\label{tab1}
\begin{ruledtabular}
\begin{tabular}{l|ccccc}
$n$ &$E_s-E_t$ (eV)&  $\tau_S$(ps) & $\tau_T$(ps) & $r$           & $\phi$ \\
\hline
 2    &  1.552                & 29.3541           &20.4905         &0.698&0.19\\
4     & 1.309                 &23.0852             &30.4756        &1.323&0.31\\
 8    & 1.199                 &18.8897              &49.8341      &2.646&0.47 \\
16   & 1.174                &17.1152              &60.9348       &3.560&0.54 \\
32   & 1.146                &17.8005              &174.771       &9.818&0.77 \\
\end{tabular}
\end{ruledtabular}

\end{table}

 Assuming that after an initial approach time, the formation 
of the lowest energy excitons follows first-order kinetics corresponding to the decay of some precurser state, we can fit the population data to an exponential form and obtain 
the formation rate constants.  For the longest chain considered ($(PPV)_{32}$),
 the formation half-times are
$\tau(S_{XT})= 17.80 $ps and $\tau(T_{XT}) =174.77 $ps, 
i.e., $r= 9.8$ corresponding to a 77\%
EL/PL efficiency, which is systematically higher than 
observed in both  LED\cite{ref3,ref4} and PADMR\cite{ref5a} data, but nevertheless consistent
with the observed trend.
For physical systems in which disorder limits the effective conjugation length, one might
expect the effective conjugation length to be more on the order of 8 to 10 repeat units. 
For PPV this corresponds to $r \approx 3$ which is more in line with the experimental observations.   


The intrinsic distinction between S and T e-h captures is readily understood
in terms of different exciton binding energies\cite{ref7,ref7b,ref8,ref8b} $\varepsilon_B=\varepsilon_{CT}-\varepsilon_{XT})$.   As seen in
Fig.\ref{DOS-PPV},  the singlet excitonic band $S_{XT}$nearly overlaps with the charge-transfer band $S_{CT}$, 
whereas in the case of the  triplet exciton, $T_1$ lies at 
the bottom of a separate band of bound e-h pairs, about 1.5eV below the
CT continuum. Thus $T_{XT}$ formation requires on average a longer sequence
of vibron-mediated transitions, each falling into the phonon frequency
range.

\subsection{Breaking e/h symmetry}
The higher efficiency of S recombination becomes more apparent when e-h
symmetry is lifted. We slightly broaden the conduction band and squeeze
the valence band, so that  $f/\overline{f}= -1.1$.  The small e-h asymmetry changes
negligibly the electronic spectrum, but opens a weak vibronic channel
for CT$\rightarrow$XT internal conversions. The resulting population dynamics are
shown in Fig.\ref{rates}b. Here we see a fast build-up of low-lying XT and CT
states in the first 100 to 200ps, followed by a slow conversion of CT
states into excitons. The initial dynamics occurs approximately without
e-h parity crossovers and the formation rates of low-lying XT and CT
singlets and triplets are about the same as in the symmetric case.
However, the triplet XT to CT branching ratio decreases drastically to
about 1:2 and suppresses the formation of $T_{XT}$. Moreover, subsequent 
$T_{CT}\rightarrow T_{XT}$ relaxation is
very slow due to the large binding energy of triplet excitons and the
low density of states between $T_{CT}$ and $T_{XT}$. 
Thus $T_{XT}$ population reaches only 40\% after
800ps propagation. In contrast, $S_{XT}$ is slightly favored over $S_{CT}$ in the
initial capture, and further $S_{CT}\rightarrow S_{XT}$ conversion occurs on a time scale
of  about  800ps.\cite{SKEB3}

Note as well that $S_{CT}$ and $T_{CT}$ are very close in energy because of
the lack of appreciable exchange between separated electrons and holes. 
Hence, intersystem $S\rightarrow T$ crossing of long-lived CT states
due to spin-orbit coupling (not included in the present model)
 is highly likely to occur prior to final e-h
binding.

\subsection{Analytic Model--Universal Scaling Relation}
Let us consider the relaxation from the lowest unbound (CT) state to the 
lowest bound  excitonic (XT) state
due to coupling to the lattice vibrations.  Since the exciton binding energy is greater than 
energy for a single phonon transition, we need to think about this a multi-phonon process.
If we assume that this occurs as a set of discrete hops of energy $\delta E \approx \hbar\omega_{phonon}$, then the binding energy is $\varepsilon_B = N \delta E$.  Furthermore, 
if we assume that the hopping rate between adjacent vibronic levels is the same for all 
steps, then the population of a given vibronic sublevel is
\begin{eqnarray}
\dot{n}_j = k (n_{i-1} - n_i)
\end{eqnarray}
where we assume that downward transitions are favored over upward transitions.  For 
systems at low temperature, this will be the case so long as $k_BT <   \delta E$.   Looking at the 
population of the state at the bottom, $n_{XT}$, which corresponds to the 
final excitonic state,  we can easily solve these coupled kinetic equations 
\begin{eqnarray}
n_{XT}(t) = 1 - e^{-kt}\sum_{i=1}^{N-2}\frac{(kt)^i}{i!}.
\end{eqnarray}
Now we ask, what is the time required for 1/2 of the initial population to appear in the lowest 
state starting from the highest?  Not surprisingly, this scales with the number of intermediate states
and consequently, the binding energy.  Thus, 
\begin{eqnarray}
r  = \frac{\tau_T}{\tau_S} \propto \frac{\varepsilon_B^T}{\varepsilon_B^S}
\end{eqnarray}
which is consistent with our simulations discussed above.  

Next, taking the singlet-triplet energy gaps into consideration.  To a lowest 
order approximation the singlet-triplet energy difference is approximately
equal to twice the electron/hole correlation energy $\Delta  E_{ST} = 2 K$.  
Using the data in Table 1,  we can fit $\Delta E_{ST}$ to a simple functional form 
\begin{eqnarray}
\Delta E_{ST} = K_\infty + K^{(1)}/n +K^{(2)}/2n^2,
\end{eqnarray}
where $K_\infty = 1.13 eV$ and $K^{(1)}=0.55 eV$ and $K^{(2)} =0.30 eV$, where $n$ is the chain length. 
Thus, 
\begin{eqnarray}
r &\propto& \frac{\varepsilon_B^T}{\varepsilon_B^S} 
= \frac{\varepsilon_B^S + \Delta E}{\varepsilon_B^S} \\
& \propto& 
\frac{\varepsilon_B^S +  K_\infty + K^{(1)}/n +K^{(2)}/n^2}{\varepsilon_B^S}
\end{eqnarray}
Finally, using $\varepsilon_B^S \propto 1/n$, we obtain
\begin{eqnarray}
r \propto (1+K^{(1)}  + n K_\infty + {\cal O}(K^{(2)}/n).
\end{eqnarray}
Since $K_\infty$  is twice the correlation energy of an electron/hole pair in 
an infinitely long chain, this we can assume that for organic polymers, its value is more or less 
system independent.   This universality consistent with the experimental observations 
reported in Ref.\cite{ref5b}
Thus, we argue that the recombination process proceeds via  a cascade through a series of 
intermediate states coupling the lowest charge-transfer states to the lowest excitonic 
state.

\begin{figure}[t]
\includegraphics[width=4.0in]{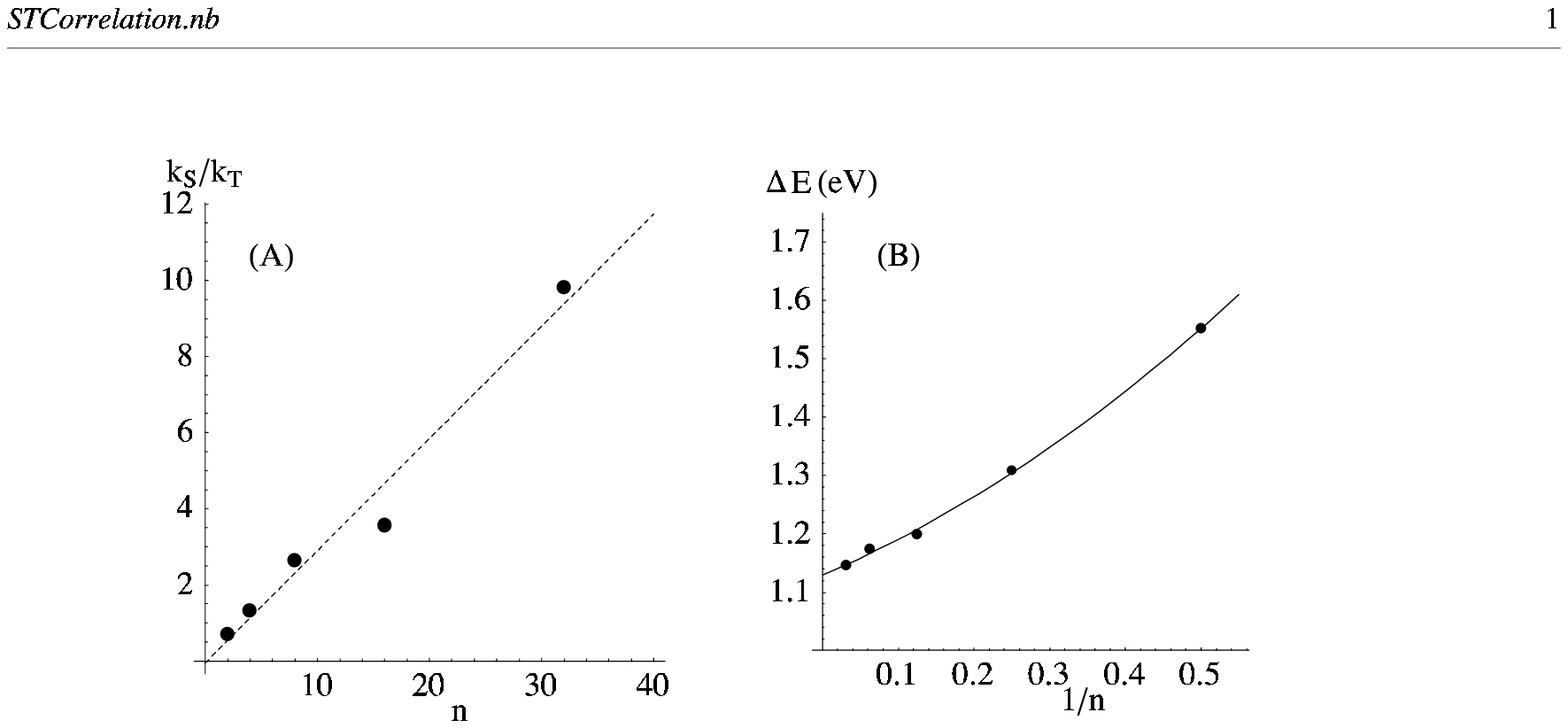}
\caption{A. Linear relation between polymer chain length, $n$ and 
the rate-constant ratios for singlet and triplet cases.
B. Singlet-Triplet gap vs. $1/n$. Solid curve is fit to quadratic form.}\label{rvn}
\end{figure}

\section{Conclusion}
In summary, we have simulated phonon-mediated intrachain e-h
recombination in LCPs. The approach allows us to examine EL quantum
yield in LCPs further case by case by combining first-principle
description with available spectroscopic data.   While we focus here on
PPV and adjust el-ph coupling to its absorption spectrum, results can be
generalized to most nondegenerate polymers.   Apart from spin-degeneracy
statistics, efficiency of singlet e-h capture outweighs the triplet one
as a natural outcome of the higher binding energy of triplet excitons.
For both S and T processes, the relaxation mechanism involves two steps
as a result of approximate e-h symmetry.  In the initial e-h capture,
both low-lying excitons and same-spin CT intermediates are formed. The
XT to CT branching ratio may vary, but is inevitably more favorable for
singlet excitons than for triplet ones simply due to energetics.
Relaxing to the higher lying singlet
exciton requires fewer elementary relaxation steps than the lower lying triplet
exciton.
Subsequent conversion of CT
states into excitons is forbidden by e-h symmetry and occurs on a much
slower time scale once this symmetry is broken.
The weak e-h asymmetry of real conjugated systems
favors again the singlet $CT\rightarrow XT$ relaxation by a factor, commensurate with
the ratio of T to S exciton binding energies.

What is desired is experimental detection of the transient kinetics we predict in out model.
Ideally, the experiment would monitor the fluorescence signal following electron-hole injection 
onto a polymer chain.  This is an extremely experiment since electro-injection typically entails
multiple polymer chains with disordered morphology.  
This is a significant lacuna in our current model and we have begun work on incorporating 
multiple polymer chains into our theory to model inter-chain e/h transfer and recombination
dynamics. 

What is also curious is that our predictions, a
and the observation that the EL efficiency is dependent upon the rate of formation of emissive 
states, then the formation rate constant should be the limiting factor in the recombination 
process following electro-injection and not the electron/hole diffusion time.   Surprisingly, the 
diffusion times for charge carriers is significantly longer than the recombination rates reported herein 
and for that matter predicted by other groups.  Moreover, since carrier diffusion is more or less independent
of spin-state (due to the fact the carriers are non-interacting), one expects that the EL efficiency to 
not depend upon the recombination process and should reflect the 1:3 ratio of singlets to triplets in the 
system.    We hope that further experimental and theoretical work will shed some additional light onto the 
electroluminescence mechanism in organic polymers.


\begin{acknowledgments}
This work was funded by the National Science Foundation and the Robert
A. Welch Foundation.  ERB wishes to thank the organizers of the 2003 Sanabel Symposium 
and Wiley, Inc. for the Wiley International Journal of Quantum Chemistry Young Investigator 
Award. 
\end{acknowledgments}


\begin{thebibliography}{10}
\bibitem{ref17} By $S_{XT}$ we mean $S_1$ and the several other thermally populated
singlets above it. Similarly for $T_{XT}$, $S_{CT}$, and $T_{CT}$.
\end{thebibliography}
\end{document}